\documentclass[acmlarge,screen]{acmart}
\usepackage{multirow}
\usepackage{csvsimple}
\usepackage{xcolor}
\usepackage{soul}
\usepackage[textsize=small]{todonotes}
\usepackage{lineno}

\def\BibTeX{{\rm B\kern-.05em{\sc i\kern-.025em b}\kern-.08emT\kern-.1667em\lower.7ex\hbox{E}\kern-.125emX}}
    
\setcopyright{acmlicensed}
\acmJournal{IMWUT}
\acmYear{2019} \acmVolume{3} \acmNumber{3} \acmArticle{88} \acmMonth{9} \acmPrice{15.00}\acmDOI{10.1145/3351246}

\begin{document}

%
\title[Modeling Personality vs. Modeling Personalidad]{Modeling Personality vs. Modeling Personalidad: In-the-wild Mobile Data Analysis in Five Countries Suggests Cultural Impact on Personality Models}







\author{Mohammed Khwaja}
\email{mohammed.khwaja@telefonica.com}
\affiliation{%
\institution{Telefonica Alpha}
\city{Barcelona}
\country{Spain}
}
\affiliation{%
\department{Brain \& Behaviour Lab, Department of Bioengineering}
\institution{Imperial College London}
\country{UK}
}

\author{Sumer S. Vaid}
\email{sumer@stanford.edu}
\affiliation{%
\department{Department of Communication}
\institution{Stanford University}
\city{Stanford}
\state{CA}
\country{USA}
}

\author{Sara Zannone}
\email{s.zannone14@imperial.ac.uk}
\affiliation{%
\department{Computational Neuroscience Lab, Department of Bioengineering}
\institution{Imperial College London}
\country{UK}
}

\author{Gabriella M. Harari}
\department{Department of Communication}
\email{gharari@stanford.edu}
\affiliation{%
\institution{Stanford University}
\city{Stanford}
\state{CA}
\country{USA}
}

\author{A. Aldo Faisal}
\email{a.faisal@imperial.ac.uk}
\affiliation{%
\department{Brain \& Behaviour Lab, Department of Bioengineering \& Department of Computing \& Data Science Institute}
\institution{Imperial College London}
\country{UK}
}

\author{Aleksandar Matic}
\email{aleksandar.matic@telefonica.com}
\affiliation{%
\institution{Telefonica Alpha}
\city{Barcelona}
\country{Spain}
}

\renewcommand{\shortauthors}{Khwaja et al.}

\begin{abstract}

Sensor data collected from smartphones provides the possibility to passively infer a user's personality traits. Such models can be used to enable technology personalization, while contributing to our substantive understanding of how human behavior manifests in daily life. A significant challenge in personality modeling involves improving the accuracy of personality inferences, however, research has yet to assess and consider the cultural impact of users' country of residence on model replicability. We collected mobile sensing data and self-reported Big Five traits from 166 participants \textcolor{black}{(54 women and 112 men)} recruited in five different countries (UK, Spain, Colombia, Peru, and Chile) for 3 weeks. We developed machine learning based personality models using culturally diverse datasets - representing different countries - and we show that such models can achieve state-of-the-art accuracy when tested in new countries, ranging from 63\% (Agreeableness) to 71\% (Extraversion) of classification accuracy. Our results indicate that using country-specific datasets can improve the classification accuracy between 3\% and 7\% for Extraversion, Agreeableness, and Conscientiousness. \textcolor{black}{We show that these findings hold regardless of gender and age balance in the dataset. Interestingly, using gender- or age- balanced datasets as well as gender-separated datasets improve trait prediction by up to 17\%.} We unpack differences in personality models across the five countries, highlight the most predictive data categories (location, noise, unlocks, accelerometer), and provide takeaways to technologists and social scientists interested in passive personality assessment. 
\end{abstract}


\maketitle

\section{Introduction}
\textcolor{black}{The use of "stable" and "underlying" characteristics to describe and explain human behavior has been an enduring concept since the nascent days of psychology~\cite{allport1937personality, cattell1957personality}}. \vphantom{Within the field of psychology}These stable behavioral characteristics are referred to as "dispositional tendencies" or personality traits \cite{goldberg1993structure, ajzen1987attitudes}. Of the numerous models that conceptualize different personality traits, psychologists have rallied around the Big Five Model, a set of psychometrically robust survey assessments that measure five dimensions of stable personality traits across \textcolor{black}{demographically} diverse individuals from different cultures~\cite{mccrae1997personality}. 

The five dimensions of personality traits characterized in the Big Five Model are: Extraversion, Agreeableness, Conscientiousness, Neuroticism and Openness. Extraversion characterizes levels of "gregariousness", "assertiveness" and \textcolor{black}{"excitement-seeking drives"}\vphantom{talkativeness}; Agreeableness captures the extent to which individuals are\vphantom{helpful, cooperative and sympathetic} \textcolor{black}{"trusting", "straightforward" and "altruistic"}; Conscientiousness captures the extent to which individuals are\vphantom{disciplined, achievement-oriented and organized} \textcolor{black}{"competent", "organized" and "dutiful"}; Neuroticism captures the extent to which individuals are \vphantom{emotionally stable}\textcolor{black}{"anxious", "irritable" and "self-conscious"}; Openness captures the\vphantom{intellectual curiosity of individuals as well as their inclination towards seeking a diversity of experiences} \textcolor{black}{extent to which individuals are "curious", "imaginative" and have "wide interests"} \cite{john2008paradigm, john1999big}.

Traditionally, personality assessments rely on survey questionnaires to quantify the five trait dimensions for each individual based on answers to a series of questions about dispositional tendencies in  affect, cognition, and behavior ~\cite{johnson2014measuring, john1999big}. However, the past decade of research has revealed that digital traces of behavior created from the use of digital media technologies (e.g., social media, smartphones) can be analyzed to infer an individual's standing on these five trait dimensions of personality \cite{kosinski2013private, vinciarelli2014survey}. 

This automatic modelling of personality traits from continuous and unobtrusive streams of data has immense potential for improving the usability and effectiveness of persuasive technologies. For instance, Halko et al. provided a set of guidelines for adapting persuasive technologies to individual personalities to improve the likelihood of success in persuading people to adopt health promoting behaviors~\cite{halko2010personality}. Moreover, Oliveira et al., found that more extroverted and conscientious individuals reported greater usability of mobile services compared to people who were less extroverted and conscientious, suggesting a link between personality traits and perception of digital products ~\cite{oliveira2013influence}.  

Given the ubiquity of smartphones, a growing body of research has extended this trait prediction literature to the realm of mobile sensing research. Mobile sensing research is focused on using the data generated from sensors in smartphones and wearable technology to infer user behavior as it unfolds in the course of day-to-day experiences ~\cite{harari2016using}. Smartphones embed a number of sensors that provide abundant behavioral data ~\cite{lane2010survey}. For example, Global Positioning Systems (GPS) capture mobility patterns, smartphone-embedded accelerometers infer the degree and duration of physical movement, Bluetooth scanning techniques detect physical proximity to other smartphones and the microphone infers the levels of sociability occurring around and by the user ~\cite{harari2017smartphone}. Previous research suggests that such data can be used to accurately classify personality traits of diverse individuals ~\cite{de2013predicting}. For instance, Chittaranjan, Blom and Gatica-Perez showed that data collected from smartphones over 8 months could be used to classify the Big Five personality type of users with accuracy equal or greater than 70 percent for all five traits ~\cite{chittaranjan2011s}. In addition, more recent research suggests that within-person variations in sensed behavioral patterns can also be used to predict personality traits ~\cite{wang2018sensing}. 

Despite initial efforts to develop personality models from smartphone data, little is known about how sensor-driven personality inferences generalize across cultural contexts (i.e., users living in different countries).  The large majority of existing work on modelling personality traits from streams of mobile sensing data has been limited to studies in one or two countries with minimal regard to cultural impact on the model accuracy and replicability outside the country where the model was developed. To realize the full potential of passive personality inference around the world, further research is needed to investigate if personality can be modeled from mobile sensing data collected from culturally diverse participants, and whether the models are transferable from one country to another.

Cultural settings within which personality\vphantom{unfolds} \textcolor{black}{processes occur}\vphantom{are} often\vphantom{highlighted}\vphantom{for their potential} impact\vphantom{on} behaviors, despite the cross-cultural nature of the Big Five model of personality ~\cite{nezlek2011cross}. Therefore, it is unclear whether a cultural impact on personality  expression would also influence the accuracy of machine learning personality trait models. In this exploratory study, we analyzed real behavior data captured using mobile sensing to examine the effect of country on sensed behaviors that are associated with personality traits. We conducted a study in five countries, namely UK, Spain, Chile, Peru and Colombia, involving a total of 166 Android users. Participating users filled out a Big Five personality questionnaire and installed a custom Android app that collected phone usage logs and sensor data for a three-week period. We extend the previous work on automatic prediction of personality traits by being the first to investigate the impact of cultural context on personality models and to explore differences in personality models across different countries. The main goal of our work is to address two research questions that remain unanswered to date:
\begin{itemize}
\item How do machine learning based personality assessment models perform across different countries? \textbf{(RQ1)};
\item What differences in the personality assessment models arise across different countries? \textbf{(RQ2)}. 
\end{itemize}

\section{Background}
\subsection{Automatic Personality Prediction}
\label{app}
The past decade has seen an increase in the literature on automatic personality recognition, especially in modelling personality traits from data generated by diverse digital and real-world behaviors \cite{vinciarelli2014survey}. For instance, a large body of research suggests that data generated from the use of social media websites such as Facebook, Twitter, and Blogging Sites can be used to accurately infer personality traits  ~\cite{hall2017say, catal2017cross, ferwerda2018you, skowron2016fusing, minamikawa2011blog}. The value of digitally derived measurements for personality assessment was exemplified in studies showing that Facebook Likes could be used to achieve higher accuracy, compared to personality scores provided by human raters ~\cite{kosinski2013private, youyou2015computer}. Similarly, researchers have shown that audio and video data, such as features of spoken conversational language and properties of in-conversation gesticulation, can be used to reliably predict the Big Five Traits. ~\cite{batrinca2012multimodal, ivanov2011recognition}. Other research has shown that gaming behavior can be used to accurately assess the personality of users ~\cite{tekofsky2013psyops}, and that the personality of a user is systematically related to specific game feature preferences ~\cite{jia2016personality}. Vinciarelli and Mohammadi~\cite{vinciarelli2014survey} provided a comprehensive literature survey on personality computing and modelling more generally.

Data generated from the use of wearable devices and smartphones has been used to model and analyze behaviors as well as predict personality trait of users ~\cite{olguin2009capturing, tong2018inference}. Smartphone sensing methods can be deployed to collect diverse data about individuals, ranging from an individual's thoughts and feelings (detected via experience sampling surveys or an analysis of language-based data) to Internet usage logs and behaviors (detected via sensor generated data and phone logs) ~\cite{harari2017smartphone}.  For example, Park et al.,~\cite{park2018simpler} relied on different data categories captured from mobile phone HTTP logs to model a set of personal characteristics including the Big Five personality traits. The study included 61 participants recruited in Spain and the authors reported the prediction accuracy between 64\% and 75\% in inferring Extraversion, Openness and Conscientiousness. In the experiments described in~\cite{chittaranjan2011s} and~\cite{chittaranjan2013mining}, Chittaranjan et al. used the phone usage logs of 83 and 117 subjects respectively over a period of 8 to 17 months in Switzerland. They analyzed the use of applications, calls, messages and Bluetooth data, reported the correlations between the extracted variables and the Big Five traits, and explored a binary classification of personality traits yielding an F-measure between 40\% and 80\% for different traits.  Using the call logs and Bluetooth scans of 53 subjects living at a University campus, the authors in~\cite{staiano2012friends} extracted social network structures represented through centrality, transitivity and triadic measures. The SVM-based binary classification resulted in the accuracy between 65\% and 80\% for recognizing the five personality traits. In a similar line, Oliveira et al.~\cite{de2011towards} relied on call logs acquired from mobile network data. The study investigated a set of nine structural characteristics of social networks for performing a regression analysis of the Big Five personality scores with a moderately high accuracy. In~\cite{de2013predicting}, the authors relied on call logs and location data of 69 participants from a US university reaching a mean accuracy of 42\% better than random prediction of the five personality traits. One of the most recent studies~\cite{monsted2018phone}, included by far, the highest number of participants - 636 - recruited at the Technical University of Denmark. The approach was mainly focused on quantifying social activities - the data consisted of calls, SMS, online networks, and physical proximity extracted through Bluetooth and GPS. The authors demonstrated the feasibility of using this data to detect Extraversion.

\subsection{Personality, Culture and Behavior}
\textcolor{black}{Past work summarizing  personality-culture research has delineated that an important focus for the field is "the consistency and validity of traits and their psychometric properties across cultures" \cite{church2016personality}}. For instance, the extant literature indicates that the structure of the Big Five model generalizes well to German, Portuguese, Hebrew, Chinese, Korean and Japanese samples \cite{mccrae1997personality}. Given that the Big Five model generalizes well to diverse cultural contexts, what is the relationship between Big Five traits and manifestations of specific behaviors across countries? 

Church et. al investigated how the Big Five traits were related to specific behavioral actions across two cultures: United States and Philippines \cite{church2007culture}. The researchers found that \textcolor{black}{ for Extraversion}\vphantom{in the domain of Extraversion}, culture-uniform behaviors were categorized by "gregariousness, positive emotions and warmth" but not by "assertiveness, activity and excitement-seeking drives". Hence, the authors concluded that only those behaviors that characterized "warm and cheerful sociability" were 
uniform representations of Extraversion across the two cultures. This implies that while there are common similarities of behavioral manifestations across cultures, there also exist important systematic differences: Americans indicated that they "socialized" more frequently and "smiled more at" strangers, whereas Filipinos indicated\vphantom{frequent socializing and smiling towards strangers where Filipinos indicated} a high frequency of "experiencing cheerful emotions" and an increased willingness to engage in interactions with\vphantom{shy or} unknown people. Similarly, behaviors intended to capture "insecurity" and "submissiveness" in the United States were found to be\vphantom{behavioral}  \textcolor{black}{typical} manifestations of Extraversion in the Philippines. \vphantom{likely accounted} \textcolor{black}{The authors attribute this effect to} the Filipino value of \textit{Pakikisama}, which encourages individuals to get along with others in order to increase their social acceptance \citep{church2008prediction}.
Cumulatively, this research suggests there are systematic similarities and differences in how personality traits manifest into behaviors across cultures.  

In a similar vein of research, Nezlek et al. investigated the link between daily social interactions and trait scores of the Big Five Model across two countries: Germany and United States~\cite{nezlek2011cross}. The researchers found similarities and differences in the relationship between social interactions and specific personality traits \textcolor{black}{across the two countries}. For instance, both countries displayed a\vphantom{robust} \textcolor{black}{positive} link between\vphantom{positive} "reactions to social interactions" and the traits Agreeableness and Conscientiousness. \textcolor{black}{Similarly, the proportion of social interactions with friends (versus others) were positively associated with the Extraversion trait in both countries.} However, there were a\vphantom{significant} number of other important differences. While there was a positive link between Extraversion and Openness \textcolor{black}{traits} with "reaction to social interactions" in the United States, this link was absent for Germany. Whereas Extraversion was related to total amount of social interaction in United States, none of the Big Five Traits \textcolor{black}{significantly} predicted the amount of social interaction in Germany. Based on these results, the authors concluded that the sociocultural context surrounding the manifestation of personality traits is an important consideration to account for in cross-cultural personality research \cite{nezlek2011cross}.

Whereas previous research suggests that the Big Five are cross-cultural traits, their behavioral manifestations seem to overlap and differ across different cultures. These systematic cultural differences in behavioral manifestations are bound to influence the effectiveness of trait inference models that have been trained on country-specific data. Thus,  further research is needed to determine how previous models predicting personality traits generalize to diverse cultural contexts. Simply put, how does one model of trait prediction developed for one specific country (i.e., United Kingdom) perform on data obtained from other countries (e.g., Spain)? 

To capitalize on the potential of using smartphone sensing to improve the effectiveness of persuasive technology and increase the quality of consumer experience, our study is the first to investigate replicability of personality models across different countries and to report on the corresponding differences in the models. The main contributions of this paper are:
\begin{itemize}
    \item Machine learning models developed from a multi-cultural data set evaluated across different countries 
    \item An analysis of differences in personality models and the most predictive data categories across different countries
    \item Takeaways for technologists and social scientists interested in passive personality assessment
\end{itemize}

In the subsequent sections, we use the terms 'culture' and 'country' interchangeably. 

\begin{figure}
  \includegraphics[width=0.7\linewidth]{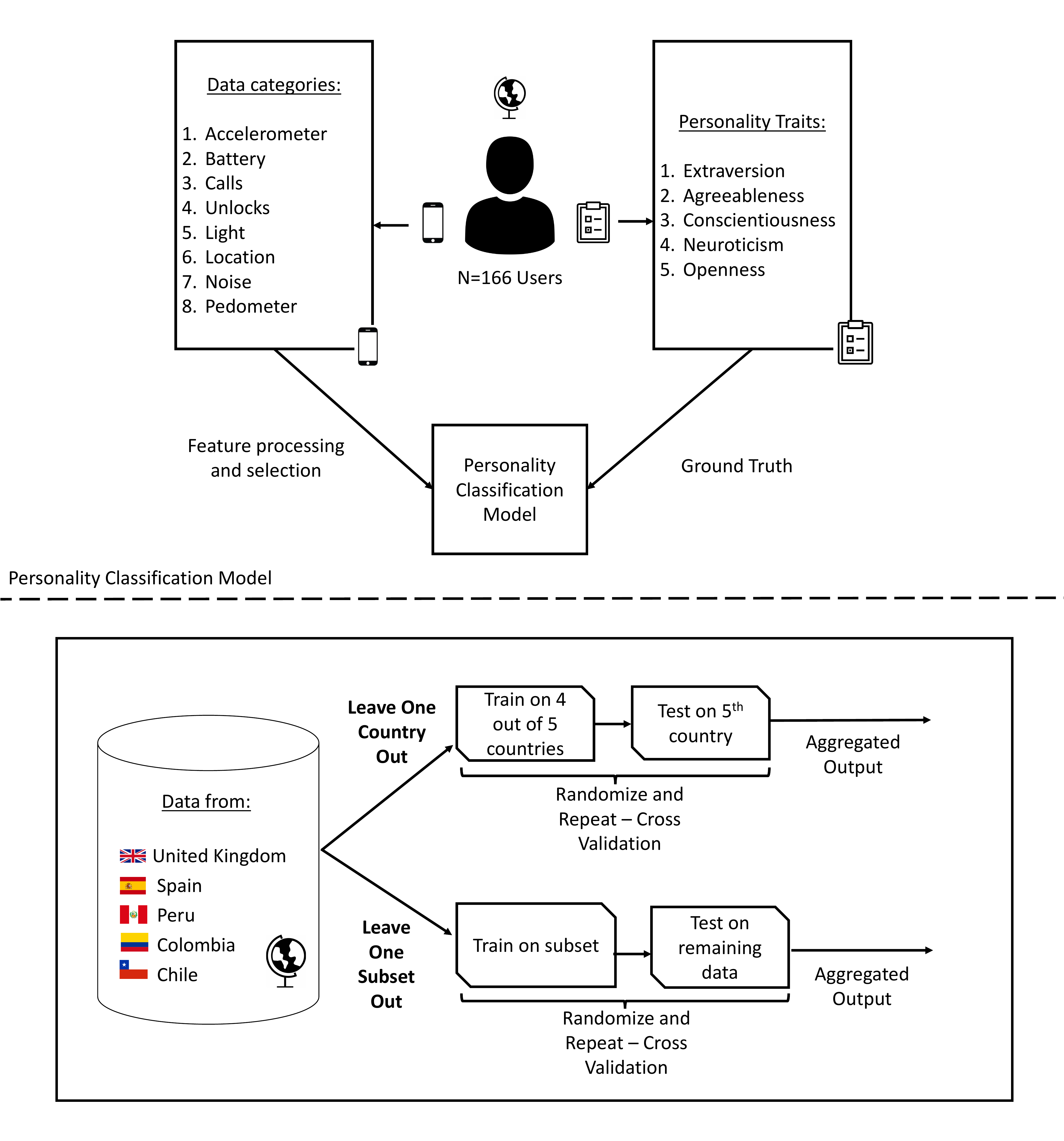}
  \caption{Flow Chart representing the methodology of this study}
  \label{fig:flow_chart}
\end{figure}

\section{Methodology}

The methodology used in this study is shown as a Flow Chart in Figure~\ref{fig:flow_chart}. Each process in this figure is described in detail in the following sections.

\subsection{Data Collection}
\label{collection}

We used a custom made Android app designed to collect data from common smartphones sensors (including light, accelerometer, pedometer, location, microphone) and to store the phone usage logs (including phone lock/unlock events, battery level, and call events\footnote{We omitted SMS logs due to a very low frequency of texting using SMS service accross all the included countries} ). The sensing modalities and features are summarized in Table~\ref{tab:data_features}. The data was collected using several types of sampling: Event based sampling was used for screen on/off events, phone lock states, and battery charging logs, accelerometer data sampling was triggered when the phone detected that a person was engaged in a physical activity (e.g., walking), whereas data from the other sensors was collected periodically by sampling from them every 15 minutes. To evaluate the energy efficiency of the data collection app, we tested it with 10 different phone models and the battery consumption stayed between 1\% and 8\%. During the study we received no complaints about increased battery consumption. 

The participants in our study were recruited through an external agency. Participants were asked to install the Android app, to keep it active for 3 weeks, and to fill out a set of on-boarding questionnaires including the 50 item Big Five Personality inventory~\cite{goldberg2006international}. In compliance with the EU General Data Protection Regulation (GDPR), participants were presented with a consent form detailing the data collected and the purpose of the study. \textcolor{black}{In general, the Android OS requires users to consent to the use of various sensor modalities by an app. Repeated permissions through pop-ups are put forth to users when the app continues to use sensor data. To simplify the use of the study software so as not to jeopardize compliance during the study, users}\vphantom{They} had the flexibility to decide which sensor information they would like recorded. On successful completion, each participant was rewarded with 40 EUR (45 USD).

\begin{table}[]
 \caption{Demographics of participants}
    \centering
    \begin{tabular}{p{2.0cm}|p{4.5cm}|p{1.0cm}|p{1.0cm}|p{1.0cm}|p{1.0cm}|p{1.0cm}|p{1.0cm}}
    \toprule
    Demographic & Particular & All & UK & Spain & Peru & Col. & Chile\\
    \midrule
    \multicolumn{2}{c|}{Size of Population} & 166 & 27 & 69 & 25 & 21 & 24 \\\hline
    Age Range & 18-25 & 30 & 6 & 19 & 1 & 2 & 2 \\
    & 26-34 & 118 & 21 & 48 & 20 & 13 & 16 \\
    & 35-44 & 18 & - & 2 & 4 & 6 & 6 \\\hline
    Gender & Female & 54 & 10 & 16 & 10 & 10 & 8\\
    & Male & 112 & 17 & 53 & 15 & 11 & 16\\\hline
    Education & No Education & 1 & - & - & - & 1 & - \\
    & Primary School & - & - & - & - & - & - \\ 
    & Secondary School & 39 & 9 & 16 & 2 & 3 & 9 \\ 
    & Technical School & 60 & 3 & 18 & 14 & 13 & 12 \\
    & Bachelor & 46 & 9 & 25 & 8 & 2 & 2 \\
    & Master & 12 & 3 & 6 & - & 2 & 1 \\
    & PhD & 2 & 2 & - & - & - & - \\
    & Other & 6 & 1 & 4 & 1 & - & - \\\hline
    Employment & Employed & 54 & 13 & 12 & 9 & 8 & 11\\
    Status & Unemployed (Job Hunting) & 18 & 3 & 9 & 3 & 2 & 1 \\ 
    & Unemployed (Not Job Hunting) & 10 & 5 & - & 1 & 1 & 3 \\ 
    & Bachelor Student & 61 & 4 & 39 & 6 & 5 & 7 \\
    & Master Student & 6 & - & 3 & 1 & 1 & 1 \\
    & Retired & 2 & - & - & 1 & 1 & - \\
    & Homemaker & 3 & - & - & - & 2 & - \\
    & Other & 12 & 2 & 4 & 4 & 1 & 1 \\
    \bottomrule
    \end{tabular}
    \label{tab:demographics}
\end{table}

\subsection{Participants and Inclusion Criteria} 
\label{participants}
Over the period of 6 months - between February and August 2018 - we recruited over 1000 potential participants out of which 545 participants (261 from Spain, 79 from Peru, 58 from Colombia, 75 from Chile and 72 from the United Kingdom) successfully finished the study according to the above mentioned requirements. \textcolor{black}{We selected a recruitment agency that has presence in multiple countries, unlike most agencies that typically operate in a single country. Our requirements included homogeneity and randomization in recruitment, equal number of men and women, and 50-70 eligible participant per country, while maximizing the number of countries. The agency had a pool of potential participants across the five countries, spanning three geographical regions (South America, Southern Europe and Northern Europe). The goal of reaching 50-70 eligible participants across all countries was successfully achieved, however the agency had more men registered in the database which resulted in an unequal ratio between men and women in the final sample (in order to reach the other criteria that we set). In addition, given the fact that there are considerable differences in patterns and frequency of phone usage in younger, compared to middle-aged (and older) populations~\cite{berenguer2017smartphones}, we opted to focus our study on age groups up to 44 years (following the age ranges from ~\cite{berenguer2017smartphones}). Due to a larger pool of participants from Spain than in other countries and perhaps an unexpectedly higher acceptance rate for participation in this country, the final sample included 250+ individuals from Spain alone (despite setting a target of 50-70 participants per country). In total, this resulted in an initial population size of 545 eligible participants across five countries.} 

We encountered a considerable portion of missing data across several data categories, which led us to reduce the analysis to the data from 166 participants who had more complete data available across the various sensing modalities. In particular, 15.6\% of Location, 9.4\% of Noise, 39.0\% of Call, 34.8\% of Accelerometer, 30.25\% of Unlock, 15.8\% of Light, and 46.9\% Pedometer data was missing. Only Battery \& Charging state logs were not missing for any participant. Location, Noise, and Call logs were primarily missing due to the fact that the participants decided to opt-out from sharing these data categories (which was the option provided in the consent form), while Pedometer, as a separate sensor, was present only in 53\% of the mobile phones in our sample of participants. Note that we relied on the Pedometer to sample the step count to avoid sampling Accelerometer data continuously in order to preserve the battery. Accelerometer, Unlock events, and Light logs were missing mostly due to technical issues, as the gaps in data occurred in an inconsistent manner - we identified particular mobile phone models and manufacturers that caused the biggest portion of the data collection gaps.

We discarded the users for whom more than 30\% of the total number of features could not be calculated due to missing data. We carefully selected the threshold of 30\% by applying the process similar to Park et al.,~\cite{park2018simpler} of exploring the model accuracy as a function of the number of participants and the data availability, choosing the threshold at the point when the model accuracy stabilizes. This reduced the number of eligible users to N=166 - Spain (N=69), Peru (N=25), Colombia (N=21), Chile (N=24) and the United Kingdom (N=27). We filled in the missing values through an iterative imputation process of computing missing feature values as a function of the non-missing features. 
The \textit{fancyimpute}\footnote{https://github.com/iskandr/fancyimpute} library was used for this purpose. 

\textcolor{black}{Given that the primary objective of our study was to explore cross-cultural machine learning models rather than the cross-cultural structural uniformity of the Big Five model, we referred to technical literature to assess the adequacy of our sample size. Note that cross-cultural personality psychology literature has typically focused on examining the structure of the Big Five traits in different settings (e.g., Schmitt et al. \cite{schmitt2007geographic}), or behavioral manifestations of traits in different settings (e.g., Church et al. \cite{church2007culture}); due to which we found that the comparison of our sample size with those used in technical literature was more appropriate. As reported in Section~\ref{app}, automatic personality inference has relied on datasets involving 48 (Wang et al. \cite{wang2014studentlife}), 53 (Staiano et al. \cite{staiano2012friends}), 69 (de Montjoye et al. \cite{de2013predicting}), 83 \& 117 (Chittaranjan et al. \cite{chittaranjan2011s, chittaranjan2013mining}) to 636 (Monsted et al. \cite{monsted2018phone}) and 646 participants (Wang et al. \cite{wang2018sensing}). Therefore, our sample size of 166 subjects is comparable to existing literature, whereas it goes beyond the state-of-the-art by including demographically diverse participants from five different cultures.}

The demographics of all the participants from the different countries is represented in Table~\ref{tab:demographics}. Overall, the gender ratio was 2:3, and within each country the ratio was roughly the same as the overall, except in Colombia where the ratio was 1:1 and in Spain where the ratio was 1:3.\vphantom{Despite our request to balance the sample of participants, the recruitment agency was unable to deliver an equal number of participants for each country and gender due to different sizes of regional databases of participants that were available across the different countries.} The age groups of the participants ranged between 18-25 (N=30), 26-34 (N=118) and 35-44 (N=18)\footnote{We avoided asking participants for their exact age, rather we only asked for the age range}, and for each country the majority of participants fall in the bracket of 26-34. Diverse sets of education level and employment status are represented in the population. To the best of our knowledge, our paper is the first in mobile sensing literature to include a diverse population in terms of country, age, education levels and employment status, as most studies in the past seem to have been confined to a particular country, university, or employment status~\cite{chittaranjan2011s, chittaranjan2013mining, wang2018sensing}.

\begin{table}[]
 \caption{Categories of Mobile Data, their Descriptions and Extracted Features}
    \centering
    \begin{tabular}{p{2.3cm}|p{4.5cm}|p{6.2cm}|p{1.3cm}}
    \toprule
    Category & Description & {Features extracted} & Num. of Features\\
    \midrule
    {Accelerometer} & {Event based. Triggered by the activity sensor and collected every 45 seconds when a person is moving. Provides 3D acceleration (x,y,z)} & {\{Mean \& Std. Dev.\} of Fourier Frequencies and Amplitudes \newline \{Mean \& Std. Dev.\} of (Kinetic) Energy of Accelereration during \{Morning, Afternoon, Evening, Night\}} & {12}\\\hline
     {Battery} & Event based. Provides battery level, charging state & Mean of charge level and number of charges & 2\\\hline
     {Calls} & Event based. Provides duration, number and direction of calls during a day & Count of Incoming, Outgoing, Missed \& Rejected calls \newline Duration of Incoming \& Outgoing calls \newline Number of Incoming, Outgoing \& Missed correspondents & {9}\\\hline
     {Unlocks} & Event based. Provides screen on/off information and phone lock state & First minute of phone unlock during \{Morning, Afternoon, Evening, Night \& Entire Day\} and last minute of phone unlock during day \newline Duration of time phone usage between phone unlock and subsequent lock \newline Interval of time between unlock and subsequent unlock \newline Count of phone unlock events & {9}\\\hline
     {Light} & Collected every 15 mins. Provides the intensity of light in lux & Median level of light during \{Morning, Afternoon, Evening, Night \& Entire Day\} & {5}\\\hline
     {Location} & Collected every 15 mins, when location is enabled. Provides GPS coordinates (Latitude and longitude) & Entropy, Count and Duration of Stay at all places, \& count of stops during day \newline Radius of Gyration \newline Time at home during \{Morning, Afternoon, Evening, Night \& Entire Day\} and time at work during entire day \newline Distance travelled and duration of time spent in travel \newline Routine Index & {13+1*}\\\hline
     {Noise} & Collected every 15 mins. Provides the level of noise in dB & Median Noise level in absolute and scaled values during \{Morning, Afternoon, Evening, Night \& Entire Day\} \newline Ratio of silence time duration vs total time duration of noise during \{Morning, Afternoon, Evening, Night \& Entire Day\} & {15}\\\hline
     {Pedometer} & Event based. Provides step count during an activity, triggered by activity sensor & Count of steps during \{Morning, Afternoon, Evening, Night \& Entire Day\} & {5}\\
    \bottomrule
    \end{tabular}
    \raggedright{\small{*Routine Index is calculated during a time period and is not counted towards the count of daily features}}
    \label{tab:data_features}
\end{table}

\subsection{Feature Extraction} 
\label{extraction}

In line with past literature ~\cite{lane2010survey} we selected eight data categories coming from sensors and phone usage logs, provided in Table~\ref{tab:data_features}. Using this data, we extracted a set of features that describe typical patterns of user behavior over the study period.

As the first step, we computed daily features for each data category as summarized in Table~\ref{tab:data_features}, following the literature in mobile sensing based personality prediction~\cite{canzian2015trajectories, chittaranjan2013mining, wang2018sensing, monsted2018phone, oliveira2013influence, de2013predicting}. For specific features, we calculated values for different parts of a day - morning from 4AM to 12PM, afternoon from 12PM to 6PM, evening from 6PM to 10PM and night from 10PM to 4AM. 
The daily features were then categorized into week days and weekend days to capture behaviors both during working days and leisure time. Location Routine Index, as defined in~\cite{canzian2015trajectories} was the only feature that was calculated for an entire period rather than daily.
Finally, we used daily features to compute mean and standard deviation (both for week days and weekend days), the former to capture the typical individuals' routines and behavior during the study period, and the latter to capture within-person variability~\cite{wang2018sensing}. In total, we obtained 284 features from 8 data categories; the number of features in each category is provided in Table~\ref{tab:data_features}.

\subsection{Feature Selection and Model Building}
\label{model}

We approach personality modeling as a machine learning classification. In this regard, we split participants into two classes, below and above the median value for each of the traits. This method is typically applied for exploring classification of personal traits~\cite{chittaranjan2011s, chittaranjan2013mining, vinciarelli2014survey, staiano2012friends}.

We initially evaluated geometrical, probabilistic and tree-based machine learning techniques, including Support Vector Machines, Naive Bayes, and Random Forest. Random Forest outperformed the other methods and it was our choice for this classification task. Following the latest literature we also tested an optimized gradient boosting technique, implemented through the \textit{XGBoost}~\cite{chen2016xgboost} library, which didn't yield improvements in performance when compared to Random Forest. As the number of users (N=166) is smaller than the number of features (284), Recursive Feature Elimination was performed in conjunction with the Random Forest method to iteratively reduce and determine the most important features to be used for a model.



In the following, we describe the process that we applied for training and testing of personality models with respect to our research questions while also taking measures to avoid overfitting. In particular, the cross-validation process was repeated a large number of times to introduce enough randomness to the models. We observed highly similar results during each iteration, whereas model tuning parameters did not impact the accuracy (which was expected for the Random Forest model).

\textbf{RQ1.} To explore the feasibility of developing personality models from culturally-diverse data sets and evaluate their applicability and accuracy in new countries (i.e. countries not included in the training set), we tested the model by applying a cross-validation technique in \textit{leave-one-country-out} manner. We sequentially excluded all instances (i.e. participants) from a single country and used these to test the model that was trained using the data from the four remaining countries. This procedure was repeated five times for each country. This process was designed to resemble a practical situation in which a personality model developed from a culturally diverse dataset is applied to predict personality in a new country. 

Due to the fact that the Spanish sample (N=69) outnumbered the other four countries (N=24$\pm$3), we sub-sampled n random participants from the Spanish data set, where n corresponds to the lowest number of samples in across five countries (in our case, N=21 in Colombia) and performed the above described cross-validation. This process was repeated 10\footnote{10 iterations was selected following the rule of thumb criteria - square root of the number of instances which was in this case 69 (the number of participants in the Spanish data set)} times, and the overall aggregated performance was obtained. Thus, in each cross-validation iteration, the total number of samples in the population was N=118 - 27 from United Kingdom, 25 from Peru, 21 from Colombia, 24 from Chile and 21 (randomly sampled from 69) from Spain. Using roughly an equal number of participants from each country allowed us to represent each country uniformly in the training and test sets.

Additionally, to understand the importance of including culture-specific data (i.e. data collected in the country where the model will be tested) in the model training process, we applied cross-validation as \textit{leave-one-subset-out}, in which n random instances i.e. participants (forming a subset) were sequentially excluded to constitute a test-set, whereas the remaining instances were used to train the model. This was done to resemble a practical situation in which a classifier was trained and tested in the same multi-cultural settings. 

Unlike the typical k-fold cross-validation, the size of each subset in the test-set corresponded to the number of participants in each country. This was done to match the test conditions in the previous model (leave-one-country-out) to allow for a fair comparison. The leave-one-set-out process was repeated 15 times to ensure a representative accuracy averaging from the random selection\footnote{As before, 15 iterations was set as a value above the square root of the number of instances, to meet the probability criteria that each sample in the set will be included in the testing set at least once}. In this analyses, we added 5 binary country flags (a dummy variable) to the other features. The data analysis, processing and classification was performed in \textit{Python} using the \textit{pandas} and \textit{scikit-learn}~\cite{pedregosa2011scikit} library. 

\textcolor{black}{As shown in Table~\ref{tab:demographics}, males outnumbered females in our sample, whereas the age range of 26-34 was a predominant age group. To understand if the results obtained from our analyses are influenced by gender, we performed the same analysis independently for the women population only, for the male population only and finally for a population that balanced the number of males and females. To obtain the balanced population, we sub-sampled n random instances from the male population, where n corresponds to the size of the female population and we repeated this process 10 times (set as the rule of thumb criteria described above). Similarly, we examined the impact of age on the personality models by performing the analysis for a balanced population of the three different age ranges. We conducted this procedure by randomly sub-sampling the population of 26-34 year old individuals to match the population size of the group with the lowest number of individuals (i.e. 35-44 year old group) and repeated it 10 times. We avoided testing the model for the three age groups separately due to a relatively small number of participants for the age groups of 18-25 and 35-44.}

\textbf{RQ2.} To explore differences in the personality models across different countries we developed a personality model (for all five traits) for each country independently and exported the most predictive features to analyze. This was performed by applying a typical leave-one-out method of sequentially excluding one instance (i.e. participant) and training the model with the remaining instances. In each step we extracted the most predictive features and the final feature importance was accumulated to form the list of the top predictive features. This process was performed to assess country specific features and not to analyze the accuracy per each country model. We intentionally opted not to explore single country models or specific demographics due to a relatively low number of participants per each country or a demographic group which would not provide solid conclusions. 

\section{Results} 

\begin{table}
  \caption{Statistics for the Big 5 Personality Scores}
  \label{tab:stats}
  \begin{tabular}{p{1.5cm}|p{1.4cm}|p{1.8cm}|p{2.0cm}|p{2.6cm}|p{1.8cm}|p{1.6cm}}
    \toprule
    Population & Statistic & \multicolumn{5}{c}{Personality Trait}\\
    \cline{3-7}
    (Size) & & {Extraversion} & {Agreeableness} & {Conscientiousness} & {Neuroticism} & {Openness}\\
    \midrule
    {Complete} & Mean & 30.01 & 39.50 & 34.17 & 29.34 & 36.81\\
    (N=166) & Std. dev. & 7.42 & 5.56 & 5.55 & 7.83 & 5.01\\
    & Median & 31.0 & 40.0 & 34.0 & 30.0 & 37.0\\
    & Max & 48.0 & 50.0 & 50.0 & 48.0 & 50.0\\
    & Min & 10.0 & 14.0 & 18.0 & 10.0 & 19.0\\\hline
    {United} & Mean & 27.72 & 39.03 & 33.65 & 27.09 & 36.03\\
    Kingdom & Std. dev. & 9.58 & 7.26 & 6.16 & 8.23 & 6.15\\
    (N=27) & Median & 28.0 & 40.0 & 33.5 & 26.0 & 36.0\\
    & Max & 48.0 & 50.0 & 48.0 & 46.0 & 49.0\\
    & Min & 10.0 & 16.0 & 21.0 & 10.0 & 25.0\\\hline
    {Spain} & Mean & 30.48 & 40.17 & 33.18 & 28.67 & 36.46\\
    (N=69) & Std. dev. & 7.37 & 5.43 & 5.45 & 7.83 & 4.79\\
    & Median & 31.0 & 40.0 & 33.0 & 29.0 & 36.0\\
    & Max & 47.0 & 50.0 & 47.0 & 48.0 & 50.0\\
    & Min & 12.0 & 14.0 & 18.0 & 10.0 & 19.0\\\hline
    {Peru} & Mean & 31.33 & 38.76 & 34.87 & 30.80 & 37.54\\
    (N=25) & Std. dev. & 5.91 & 4.67 & 5.67 & 6.36 & 4.26\\
    & Median & 32.0 & 39.0 & 34.0 & 31.0 & 37.0\\
    & Max & 43.0 & 48.0 & 47.0 & 47.0 & 49.0\\
    & Min & 17.0 & 25.0 & 22.0 & 15.0 & 26.0\\\hline
    {Colombia} & Mean & 29.84 & 38.37 & 36.60 & 32.70 & 38.5\\
    (N=21) & Std. dev. & 6.75 & 4.58 & 5.00 & 7.51 & 4.62\\
    & Median & 30.0 & 39.0 & 36.5 & 34.5 & 38.0\\
    & Max & 45.0 & 47.0 & 50.0 & 45.0 & 48.0\\
    & Min & 15.0 & 21.0 & 27.0 & 17.0 & 22.0\\\hline
    {Chile} & Mean & 29.22 & 39.14 & 35.31 & 29.56 & 36.62\\
    (N=24) & Std. dev. & 6.83 & 5.58 & 4.71 & 8.23 & 5.31\\
    & Median & 30.0 & 39.0 & 35.0 & 30.0 & 37.0\\
    & Max & 43.0 & 49.0 & 48.0 & 45.0 & 49.0\\
    & Min & 11.0 & 24.0 & 20.0 & 10.0 & 22.0\\
  \bottomrule
\end{tabular}
\end{table}

\subsection{Questionnaire Analysis} 
The Big Five Personality inventory~\cite{goldberg2006international} consists of a set of 50 questions rated on a Likert scale from 1-5, to compute scores for five personality types: Extraversion, Agreeableness, Conscientiousness, Neuroticism and Openness. The minimal and maximal values for each trait are 10 and 50, respectively. Table~\ref{tab:stats} shows the statistics from the five countries in our study, including the mean, standard deviation, median, maximum and minimum. There are similar to those reported in previous work~\cite{gurven2013universal, staiano2012friends}. The personality scores in our study showed a good internal reliability, with Cronbach's alpha of $\alpha > 0.7$), with the highest $\alpha$ values observed for Neuroticism ($\alpha = 0.89$) and Extraversion ($\alpha = 0.86$), similar to the previous work~\cite{gow2005goldberg}. We observed that all the traits were normally distributed (by applying Shapiro-Wilk normality test), as well as that each trait was normally distributed within a single country ($p>0.05$). Therefore, we used the median value for each trait to split participants into the two groups for classification, as performed in previous literature~\cite{vinciarelli2014survey}, which resulted in balanced classes in terms of the number of participants in each class (with the baseline accuracy always being close to 50\%). 


\subsection{Personality Inference} 

\begin{table}
  \caption{Results obtained from classification analysis}
  \label{tab:results_classification}
  \begin{tabular}{p{1.55cm}|p{1.45cm}|p{0.85cm}|p{0.85cm}|p{0.85cm}|p{0.85cm}|p{0.85cm}|p{0.85cm}|p{0.85cm}|p{0.85cm}|p{0.85cm}|p{0.85cm}}
    \toprule
    Population & Method & \multicolumn{10}{c}{Personality Trait}\\
    \cline{3-12}
     & & \multicolumn{2}{c|}{Extraversion} & \multicolumn{2}{c|}{Agreeableness} & \multicolumn{2}{c|}{Consc.} & \multicolumn{2}{c|}{Neuroticism} & \multicolumn{2}{c}{Openness}\\
     \cline{3-12}
     & & Acc (\%) & $\kappa$ & Acc (\%) & $\kappa$ & Acc (\%) & $\kappa$ & Acc (\%) & $\kappa$ & Acc (\%) & $\kappa$\\
     \midrule
     \multirow{8}{*}{Complete} & Method 1: leave-one-country-out & 71 & 0.41 & 63 & 0.29 & 68 & 0.35 & 72 & 0.42 & 70 & 0.36\\
     \cline{2-12}
     & \rule{0pt}{3ex}Method 2: leave-one-subset-out & 74 & 0.48 & 70 & 0.37 & 71 & 0.39 & 72 & 0.44 & 69 & 0.35\\
  \bottomrule
\end{tabular}
\end{table}

Table~\ref{tab:results_classification} presents the performance of the two personality classification models  - one developed by using leave-one-country-out method (Method 1) and the other by using the leave-out-subset-out method (Method 2). By design, the former method did not include data from the test country in the training set as opposed to the latter method. 

Both models demonstrated a solid performance with the accuracy significantly above the baseline, in line with the previous literature~\cite{chittaranjan2011s, chittaranjan2013mining, wang2018sensing}. The accuracy of the model that included the \textit{knowledge} i.e. the data from the country where it will be tested was ranging between 69\% and 74\%. On the other hand, the accuracy of the model that did not have any prior knowledge about the test country performed between 63\% and 71\%. Table~\ref{tab:results_classification} also includes Cohen's $\kappa$ values that are indicative of how well the classifier performs with respect to the baseline of classifying all instances in a single class - $\kappa = 0$ indicates baseline performance corresponding to a classification accuracy of 50\%, as the data is equally distributed around the median. 

To compare the accuracy of the two models (Method 1 and Method 2), we relied on the McNemar's test. The test indicated a significant statistical difference for Extraversion ($p<0.001$), Agreeableness ($p<0.001$) and Conscientiousness ($p<0.05$), and no statistically significant difference for predicting Neuroticism and Openness. The results of  McNemar's test show that having culture-specific dataset in the training data-set significantly improves the prediction accuracy for Extraversion, Agreeableness and Conscientiousness, whereas the model of Neuroticism and Openness appears to be more culture robust. Further improvements of the model (Method 2) may be possible (e.g. by putting more weight to the instances from the country where the model is applied) however it is out of the scope of this paper. Importantly, we showed that using multi-cultural data-sets to develop the personality models can predict personality in a new country (not used to train the classifier).

{\color{black}\subsection{Effect of Demographics}

The personality classification models for different gender populations are presented in Table~\ref{tab:results_gender}. As described previously, Method 1 refers to the leave-one-country-out (i.e. testing a country-agnostic model) and Method 2 refers to the leave-one-subset-out method (i.e. testing a country-specific model). When balancing the number of male and female participants in the dataset, the accuracy of the country-specific model (i.e. Method 2) ranged between 74\% and 81\%, while for the country-agnostic model the accuracy ranged between 73\% and 75\%. In comparison to the results that relied on the full dataset (without achieving the gender balance, as reported in Table~\ref{tab:results_classification}), a balanced dataset of male and female achieves a higher accuracy for both Method 1 and Method 2. McNemar's test to compare Method 1 and 2 indicated significant statistical difference for Extraversion ($p<0.01$), Agreeableness ($p<0.001$), Conscientiousness ($p<0.001$) and Neuroticism ($p<0.001$). This means that our findings indicating a higher predictive power of country-specific than country-agnostic models hold irrespective of gender.

\begin{table}
  \caption{Results obtained from classification analysis from populations of different genders}
  \label{tab:results_gender}
  \begin{tabular}{p{1.55cm}|p{1.45cm}|p{0.85cm}|p{0.85cm}|p{0.85cm}|p{0.85cm}|p{0.85cm}|p{0.85cm}|p{0.85cm}|p{0.85cm}|p{0.85cm}|p{0.85cm}}
    \toprule
    Population & Method & \multicolumn{10}{c}{Personality Trait}\\
    \cline{3-12}
     & & \multicolumn{2}{c|}{Extraversion} & \multicolumn{2}{c|}{Agreeableness} & \multicolumn{2}{c|}{Consc.} & \multicolumn{2}{c|}{Neuroticism} & \multicolumn{2}{c}{Openness}\\
     \cline{3-12}
     & & Acc (\%) & $\kappa$ & Acc (\%) & $\kappa$ & Acc (\%) & $\kappa$ & Acc (\%) & $\kappa$ & Acc (\%) & $\kappa$\\
     \midrule
     \multirow{2}{*}{Balanced} & Method 1 & 73 & 0.46 & 75 & 0.51 & 73 & 0.46 & 74 & 0.47 & 75 & 0.47\\
     \cline{2-12}
     & \rule{0pt}{3ex}Method 2 & 76 & 0.53 & 81 & 0.57 & 77 & 0.52 & 78 & 0.55 & 74 & 0.46\\
     \cline{1-12}
     \rule{0pt}{3ex}\multirow{2}{*}{Female} & Method 1 & 78 & 0.55 & 81 & 0.61 & 78 & 0.60 & 76 & 0.51 & 75 & 0.51\\
     \cline{2-12}
     & \rule{0pt}{3ex}Method 2 & 80 & 0.61 & 87 & 0.73 & 79 & 0.61 & 78 & 0.56 & 74 & 0.46\\
     \cline{1-12}
     \rule{0pt}{3ex}\multirow{2}{*}{Male} & Method 1 & 75 & 0.51 & 68 & 0.34 & 75 & 0.44 & 76 & 0.51 & 74 & 0.46\\
     \cline{2-12}
     & \rule{0pt}{3ex}Method 2 & 79 & 0.58 & 73 & 0.47 & 77 & 0.51 & 77 & 0.52 & 76 & 0.51\\
  \bottomrule
\end{tabular}
\end{table}

Additionally, as indicated in Table~\ref{tab:results_gender}, developing models independently for males and independently for females further improves the accuracy in comparison to the full dataset as well in comparison to the gender balanced dataset. For the female population, the accuracy of the country-specific model (Method 2) ranged between 74\% and 87\%, whereas the country-agnostic model (Method 1) ranged between 75\% and 81\%. For the male population, the accuracy of the country-specific model ranged between 73\% and 79\%, while the country-agnostic model ranged between 68\% and 76\%. Consistent with the results obtained using the full dataset,  culture-specific models outperform culture-agnostic models for gender-balanced and gender-independent models, but produce better performance in accuracy for prediction of personality.
}

{\color{black} Table~\ref{tab:results_age} shows the personality classification models for a balanced population consisting of a comparable sample size for each age group. The accuracy of the country-specific model ranged between 73\% and 78\%, whereas the country-agnostic model ranged between 72\% and 75\%. McNemar's test for comparison between the two models indicated significant statistical difference for Extraversion ($p<0.05$), Agreeableness ($p<0.001$) and Conscientiousness ($p<0.001$). Also in this case, the findings related to the comparison between country-specific and country-agnostic models fully hold also when balancing the age groups available in our sample. 

In the following, we analyze the most predictive features and we also provide an in-depth discussion of the results related to the cultural differences in the models obtained from the full dataset.

\begin{table}
  \caption{Results obtained from classification analysis from a balanced population of different age ranges}
  \label{tab:results_age}
  \begin{tabular}{p{1.55cm}|p{1.45cm}|p{0.85cm}|p{0.85cm}|p{0.85cm}|p{0.85cm}|p{0.85cm}|p{0.85cm}|p{0.85cm}|p{0.85cm}|p{0.85cm}|p{0.85cm}}
    \toprule
    Population & Method & \multicolumn{10}{c}{Personality Trait}\\
    \cline{3-12}
     & & \multicolumn{2}{c|}{Extraversion} & \multicolumn{2}{c|}{Agreeableness} & \multicolumn{2}{c|}{Consc.} & \multicolumn{2}{c|}{Neuroticism} & \multicolumn{2}{c}{Openness}\\
     \cline{3-12}
     & & Acc (\%) & $\kappa$ & Acc (\%) & $\kappa$ & Acc (\%) & $\kappa$ & Acc (\%) & $\kappa$ & Acc (\%) & $\kappa$\\
     \midrule
     \multirow{2}{*}{Balanced} & Method 1 & 75 & 0.50 & 72 & 0.44 & 72 & 0.34 & 74 & 0.48 & 73 & 0.44\\
     \cline{2-12}
     & \rule{0pt}{3ex}Method 2 & 77 & 0.54 & 78 & 0.55 & 77 & 0.54 & 75 & 0.48 & 73 & 0.45\\
  \bottomrule
\end{tabular}
\end{table}}

\begin{figure}
  \includegraphics[width=0.6\linewidth]{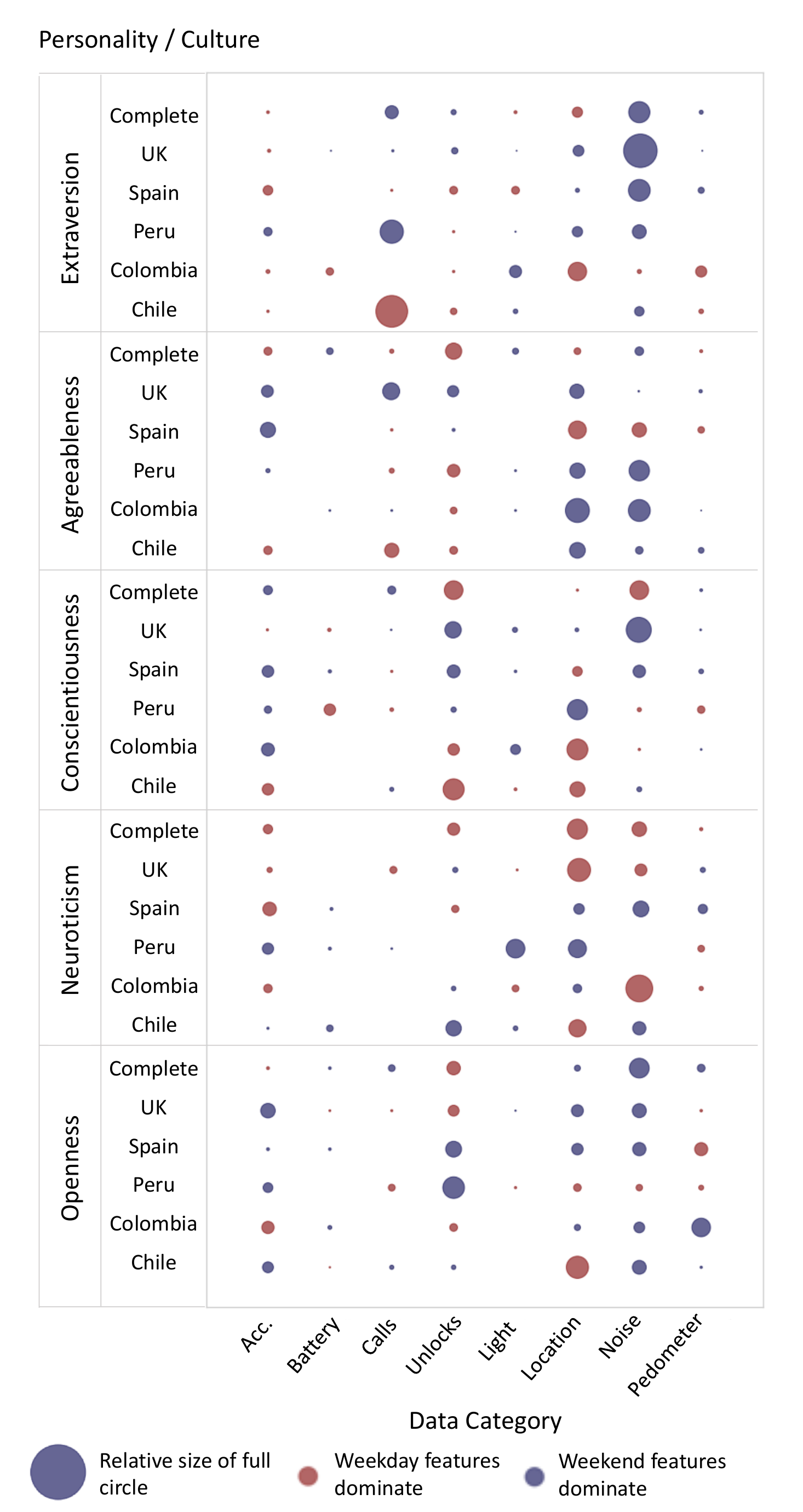}
  \caption{Top Predictive Data Categories for each Personality Trait within each dataset}
  \label{fig:data_categories}
\end{figure}

\subsection{Feature Analysis}

To obtain the top predictive features for each culture, individual models were trained and tested using the dataset from each culture, in a consistent manner to the previous model development (cross-validation of Random-Forest model implemented as leave-one-instance-out, and by applying a Recursive Feature Elimination). As previously explained, the weights assigned by the model to each feature are aggregated over each data category (provided in Table~\ref{tab:data_features}), and were used to compute the relative importance of feature categories for each trait, as represented in Figure~\ref{fig:data_categories}. In Figure~\ref{fig:data_categories}, each row corresponds to a model trained for a particular Personality trait, for the entire data-set (all five countries as per Method 2) or for a specific country (as specified). Bubbles in this figure represent the aggregate contribution for each of the 8 data categories to the model. The size of each bubble is proportional to the weight contributed by that data category to the model. The sum of all weights in a row equals 1 and this would correspond to the largest bubble possible in the figure, shown in the legend of the figure. Hence, larger bubbles represent a stronger contribution to the model by features from that data category and vice versa. Red bubbles denote the dominance of week days aggregated features from that category, and blue bubbles represent the same for weekend days. In the following, we summarize the results for each personality trait.

\textbf{Extraversion} was best predicted by Noise, Calls and Location features. Noise features were especially predictive for Spain and UK samples whereas Call features were especially predictive for Chile and Peru. None of the features were especially predictive for Colombia but Location features were the most predictive for this country. 

\textbf{Agreeableness} was best predicted by Location features, making substantial contributions in prediction across all countries - especially for Spain, Colombia and Chile. Similarly, Accelerometer data contributed moderately to Spain and UK predictions.

\textbf{Conscientiousness} was best predicted by Unlock features, making substantial contributions in predictions across all countries with the exception of Peru. Peru predictions were especially accounted for by the Location features. 

\textbf{Neuroticism} was best predicted by Location features for all countries with the exception of Colombia. Similarly, Noise features provided the most predictive power for all countries except Peru. As an exception to the rule, the otherwise non-predictive Light features contributed especially well to Peru features. 

\textbf{Openness} was best predicted by Noise, Location, Unlocks and Accelerometer based features. Noise features provided the most predictive power for the overall samples and contributed significantly to all countries. Unlock features contributed well to the Spain and Peru models, while Location was significant for prediction for Chile.

\textcolor{black}{Figure 3 shows the normalized frequency distributions of the top predictive features per country. For location, time spent at work during weekdays is positively skewed, though with considerably country-wide variation for each level of time. The distribution of time spent at home during weekend days is more uniform over time as compared to time spent at work during week days. Country-wide variability in normalized frequency of time spent at home is marginal - it is not as high as the country-wide variability of time spent at work. For noise levels at week-day nights, the distribution is somewhat bimodal, and the figure indicates that the United Kingdom and Spain are decidedly more quiet than the South American countries. The distribution of silence to noise time ratio on weekday nights also shows similar trends, with Peru, Colombia and Chile having the lowest ratios. Similarly, the distribution of weekday unlocks is slightly positively skewed and somewhat bimodal, with modest amounts of country-wide variability at each frequency level of unlocks. The unlock frequency distribution of United Kingdom and Spain is more uniform over different frequency levels as compared to the South American countries, which exhibit greater variability. Distributions of accelerometer do not appear to follow any common distribution. In general, similarities in distributions between the UK and Spain, and between Peru, Colombia and Chile indicate similarities within Europe and South America respectively.}  




\begin{figure}[htb]
  \includegraphics[width=\linewidth]{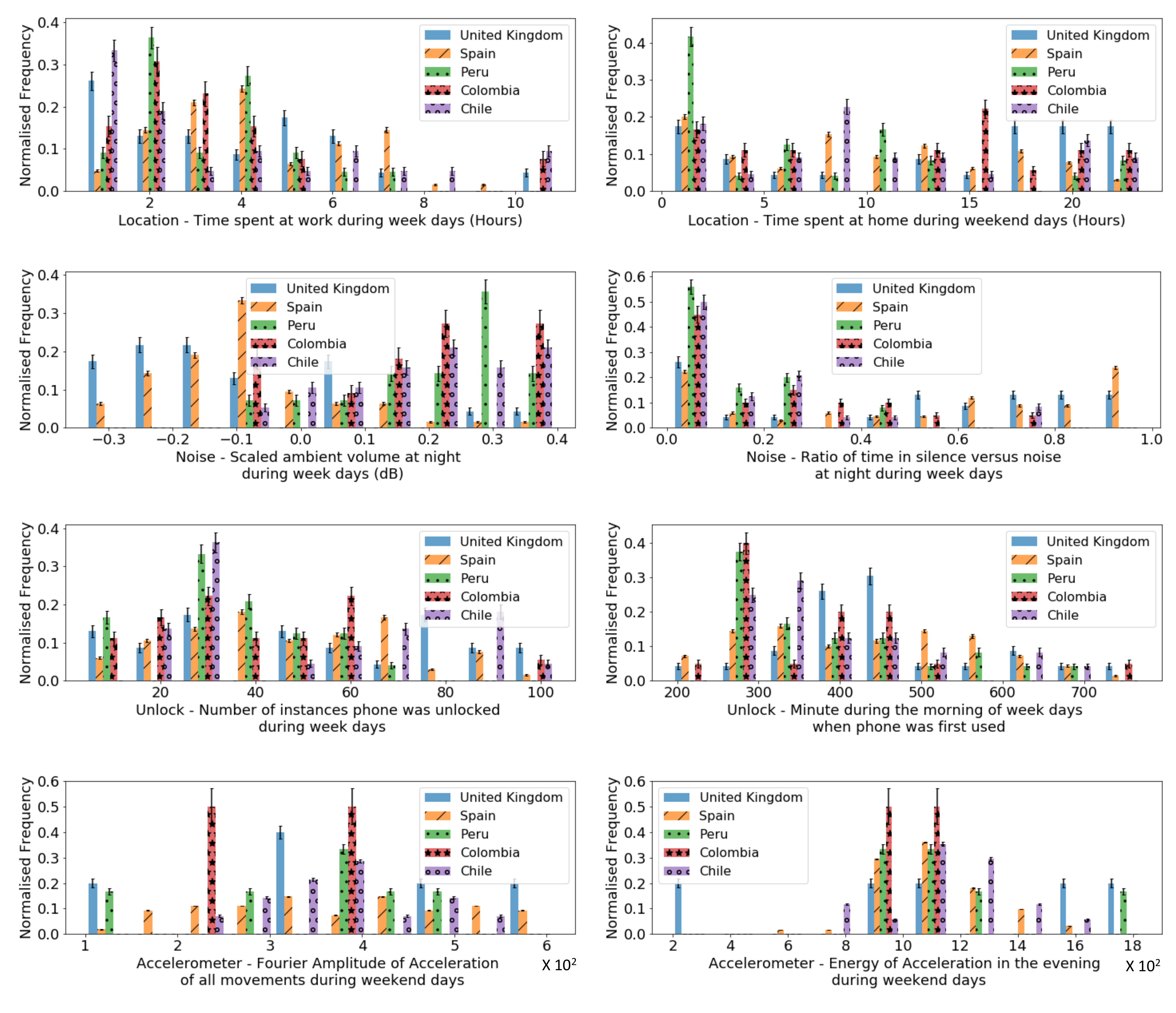}
  \caption{Normalized Frequency distributions of top predictive features of most significant data categories per country}
  \label{fig:top_feature_density}
\end{figure}

\section{Discussion} 



Here we reported results from a cross-cultural study of passive personality modeling conducted in five countries: Spain, Chile, Colombia, UK and Peru. Specifically, we developed machine-learning models to classify individuals on the Big Five trait dimensions using smartphone sensor data. Our main goal was to explore how machine learning based models perform across different countries \textbf{(RQ1)} and to examine differences in personality models that arise across different countries \textbf{(RQ2)}. 

\subsection{Personality Models in Multi-Cultural Settings \textbf{(RQ1)}}

The main findings related to the personality model development acquired from the in-the-wild mobile sensing study with 166 participants in five countries are three-fold:

\begin{itemize}
\item It is feasible to create a country-agnostic machine learning model to predict Big Five Personality Traits.

\item To achieve significant accuracy improvements in predicting Extraversion, Agreeableness and Conscientiousness, the personality models need to include a country-specific dataset, \textcolor{black}{irrespective of demographics (gender, age).}

\item \textcolor{black}{Using gender-balanced, age-balanced or gender-independent models improves the performance of personality prediction models}
\end{itemize}

The models for predicting personality traits trained in multiple countries and tested in a new country (unknown to the classifier) provided the binary classification accuracy of 71\% for Extraversion, 63\% for Agreeableness, 68\% for Conscientiousness, 71\% for Neuroticism and 70\% for Openness. Statistically significant improvements of 3\%, 7\%, 3\% in predicting respectively Extraversion, Agreeableness and Conscientiousness were achieved by including country-specific datasets. When it comes to predicting Neuroticism and Openness, country-specific data sets did not yield accuracy improvements.

\textcolor{black}{On the other hand, the accuracy of the models improved between 2\% and 17\% when using a gender-balanced, age-balanced or gender-independent dataset. Using country-specific datasets for a gender-balanced population statistically improved the prediction of Extraversion, Agreeableness, Conscientiousness and Neuroticism by 3\%, 6\%, 4\%, 4\%, but did not provide an improvement for Openness. Previous work on gender differences in the manifestation of the Big Five traits suggests that there are systematic trait and facet-level differences across gender \citep{costa2001gender}. Specifically, women have a tendency to have greater scores in Neuroticism, Agreeableness,  the "Warmth" facet of Extraversion and the "Openness to Feelings" facet of Openness \citep{costa2001gender}. Similarly, the authors found that men had a tendency to report higher scores in the "Assertiveness" facet of Extraversion and the "Openness to Ideas" facet of Openness \citep{costa2001gender}. The improvement in personality prediction for a gender-balanced and -independent population can be attributed to this difference in trait and facet level manifestation. The authors further report that personality-based gender differences were greatest for Western (European and American) cultures, as compared to other cultures \citep{costa2001gender}. Cumulatively, this research suggests that gender differences not only exist in personality trait and facet score, but are often modulated by the cultural setting within which they occur \citep{costa2001gender}. This may also indicate why the models with country knowledge improve significantly for gender-balanced and -independent models. For an age-balanced population, statistically significant improvements in prediction of Extraversion, Agreeableness and Conscientiousness by 2\%, 6\%, 5\% were observed. Furthermore, we tested the addition of quantitative labels to all the models, that describe qualitative information about users, including education levels and employment status. We observed that this addition had little to no effect on the results.}

Across all countries and all traits, we found that Noise, Location, Unlock and Accelerometer features carry the most predictive power, respectively. In each of these four groups, we identified features that were the most predictive across all the five traits. In the Location group, the most predictive features included "Time spent at home during weekend days" and "Time spent at work during week days," which were important for predicting Neuroticism. The leading features in the Noise group were "Scaled ambient volume at night during week days" and "Ratio of time in silence versus noise at night during week days," which were important for predicting Extraversion, Conscientiousness, Neuroticsm and Openness. Regarding the Unlock features, "Number unlocks during week days" and "First morning phone usage during week days" carried most of the weight, which were important for predicting Conscientiousness and Agreeableness. As for the Accelerometer features, the "Amplitude of acceleration during weekend days" and "Energy of Acceleration in the evening during weekend days" provided the most predictive power, which were important for predicting Conscientiousness.


The features in the four groups also showed different descriptive distributions across the five countries, \textcolor{black}{as shown in Figure}~\ref{fig:top_feature_density}. This suggests variations in behavioral patterns across different countries, particularly in those that turned out to be predictive of personality traits. Though these findings suggest a potential cultural impact of country on the behavioral expression of personality in mobile sensing data, we refrain from deriving substantive conclusions due to relatively small sample sets in each individual country. Interestingly, machine learning models were able to capture culture-independent similarities within same personality traits to yield a state-of-the-art accuracy level. The variations in descriptive distributions in the top predictive features may help to explain the accuracy difference between the personality models when including or excluding country specific data. 



Location and noise features provide situational information about participants' surrounding during the course of the study. Location features capture mobility patterns whereas Noise features capture properties of the environment. In contrast to describing situations, the Accelerometer and Unlocks features predominantly capture information about a person - for instance, more neurotic individuals might be checking their phone frequently while waiting for a message or before arriving at an event. 

Interestingly, the most predictive features for inferring personality traits across a wide range of countries consist of a mix of situational and personal features. Another notable observation is the higher predictive power of features extracted during Weekends as compared to Weekday features. This may be due to the fact that individual differences are captured better from activities over which individuals have more control and typically there is more choice during weekends than during weekdays.

Previous work~\cite{chittaranjan2011s, chittaranjan2013mining} has extensively used Call data for inferring personality traits, showing its predictive power especially for Agreeableness and Extraversion. In our models, however, Call features did not turn out to be predictive for these sociability-relevant traits. This might be due to a recent and considerable shift to Internet based call and messenger services such as FaceTime, WhatsApp, Viber and similar mobile apps\footnote{https://www.businessinsider.com/uk-chat-apps-overtake-calls-2018-8?IR=T}.

\textcolor{black}{Past work examining the cross-cultural stability of traits~\cite{mccrae2002cross} suggested that it is unlikely to have a single set of
questionnaire items that would be optimal in every culture. Yet, the Big Five traits were used as a universal tool in a wide variety of cultures. 
Therefore, it is reasonable to expect that smartphone based Big Five trait prediction models generated in one country would perhaps not be as predictive when applied to the data obtained from other countries. This is because not only do the behavioral manifestations of the Big Five traits vary across cultures \citep{nezlek2011cross}, but also because the very structure of the Big Five and its associated constructs might differ across different cultures \citep{mccrae2002cross}. Hence, our findings that different behavioral features derived from smartphone sensing data are differently predictive of Big Five Traits across the five countries in our sample is consistent with findings about the universality and behavioral manifestations of the Big Five traits in cross-cultural psychology.}


\subsection{Differences in Personality Models Across Countries \textbf{(RQ2)}}

There are systematic differences across countries in terms of which features confer the most predictive power for a specific Big Five Trait. Below we discuss which sensing data categories contained the most significant features to predict each trait in each country. 

For Spain, Noise features were the most predictive for four out of the five traits. The exception was for Agreeableness, which was predicted better by Location features. An example of a highly predictive feature was the "Scaled ambient volume at morning during weekend days", which was the most predictive feature for Extraversion in the Spanish population.

For the UK, Noise features were the most predictive for all traits except Agreeableness, which was predicted better by Location and Unlock features. An example of a highly predictive feature was the "Time duration of travel during week days", which was the most predictive feature for Neuroticism in the English population. 

Peru displayed the most heterogeneous set of predictive features across all the traits. Extraversion was best predicted by Call features, Agreeableness was best predicted by Location and Noise features, Conscientiousness by Location features, Neuroticism by Light and Location features whereas Openness was best predicted by Unlock features. An example of a highly predictive feature was the "Count of outgoing calls during the weekend days", which was the most predictive feature for Extraversion in the Peruvian population. 

Colombia also showed a somewhat heterogeneous mix of highly predictive features across personality traits. Location based features contributed most to the prediction of Extraversion. Similarly, Location and Noise based features contributed the most predictive power to the Agreeableness trait. Location based features offered the most predictive power to Conscientiousness, whereas Neuroticism was best predicted by the Noise features. Lastly, Openness was best predicted by Pedometer features. An example of a highly predictive feature was the "Distance travelled during weekend days", which was the most predictive feature for Agreeableness in the Colombian population. 

For Chile, Extraversion was best predicted by Call features. Agreeableness was best predicted by Location features. Conscientiousness was best predicted by Unlock features, whereas Neuroticism was best predicted by Unlock features and Location features. Lastly, Openness was best predicted by Location features. An example of a highly predictive feature was the "Time duration of outgoing calls during weekend days", which was the most predictive feature for Extraversion in the Chilean population.

\section{Limitations}

Our study had several limitations that need to be addressed in future work. \textcolor{black}{First, we relied on an external agency to sample subjects across different countries. Despite our request to balance the number of participants for each country and gender, the agency was unable to entirely meet our requirements, which resulted in a imperfectly balanced dataset. 
We also restricted the recruitment to participants younger than 44 years, as the usage of smartphones is significantly different for older populations~\cite{berenguer2017smartphones}. However, our samples across various countries were mostly homogeneous and therefore comparable (Table 1). Also, by applying a multiple random sub-sampling from the population, we tested the impact of gender and age, and we showed that our findings are robust to demographic differences.}

\textcolor{black}{Previous work in sensing literature, including~\cite{wang2018sensing, wang2014studentlife, de2013predicting, staiano2012friends} have often relied on recruiting from a student population from North America, which may have limited the external validity of the findings. In this regard, our dataset expands the literature by providing a more diverse population that includes a large portion of individuals who are not students. This prompted us to explore the differences between student and non-student populations when building personality models, given the expected differences in behavioral routines . The accuracy of country-specific models are shown in Table~\ref{tab:results_students}. The table shows that there is a difference in prediction accuracy between the two populations. 
We do not delve into these differences further as it is out of scope of this paper. However, we hope that this will serve as a call to arms for studies in personality prediction to include more demographically heterogeneous populations in the future.}

While our sample is more diverse than traditional smartphone sensing studies, our per country samples do not exceed N=21. This low sample size impairs the external validity of our findings and makes it difficult to reject data artifact-related explanations for the trend of findings observed in our data. That is, it is possible that our low sample sizes constricted our findings to a small set of country-specific users, the smartphones usage patterns of which do not generalize to the general population of individuals across cultures. This issue is further magnified by uneven sampling across different countries. More specifically, our data contained 69 Spanish subjects, while most other countries had samples in the vicinity of 21 subjects. We mitigated potential effects arising from this uneven cell size by sampling randomly from the uneven cell frequency group (i.e Spain) while training and testing the prediction accuracy of our model. Moreover, for each of the countries, our data captured diverse participants from somewhat diverse demographics (Table~\ref{tab:demographics}), providing some feasibility-related conclusions about developing country-specific prediction models. 

Importantly, in order to assess accuracy improvements when using country-specific datasets we refrained from reporting accuracy of personality models developed from each country independently due to a low per-country sample size (despite the fact that the accuracy reached 85\% in our tests). Instead, we exploited the potential of the full data set of N=166 to assess the importance of including country-specific data in the training set and to compare these to country-agnostic models.  

\textcolor{black}{Reducing our sample from N=545 to N=166 primarily occurred due to frequent opt-outs as well as technical issues that resulted in gaps across various data categories. Newer smartphones and mobile operating systems might bring more stable sensor data streams. In addition to resolving data gaps, our study might have benefited from a greater diversity in data categories, e.g. Bluetooth logs, Wi-Fi, keystroke patterns, Internet logs - that some previous studies~\cite{park2018simpler, chittaranjan2011s, roy2019probability} utilized to predict personality. However, we expect that the addition of these data categories would bring only marginal improvements, given that our models outperform the aforementioned studies. Additional data categories for personality prediction might also include the front-facing camera of the smartphone, as the link between photos and the Big Five traits has been established by literature~\cite{ferwerda2018predicting}. Also, music listening patterns can be used for personality inference, as personality was linked with music taxonomy preferences~\cite{ferwerda2015personality}. Besides, collecting very diverse and large amounts of user data (in particular, audio and video) has shown to raise privacy concerns\footnote{http://www.privacy-regulation.eu/en/article-5-principles-relating-to-processing-of-personal-data-GDPR.htm}. The EU General Data Protection Regulation encourages data minimization i.e. collecting data only from minimally needed sources and during necessary time periods. For instance, a recent study that advocated for data minimization in personality prediction has demonstrated a fairly high personality inference accuracy from smartphone data gathered only during weekends~\cite{khwaja2019interact}.}

Our dataset was collected over a period of three weeks and it is well-known that there are seasonal effects on behaviors~\cite{foster2008human} which our dataset does not capture at an individual level. However, the dataset captures seasonal effects at a sample level considering the fact that the period between February and August includes all four seasons when including the data from the northern and southern hemisphere. The personality models remained robust to the seasonal effects.

\begin{table}
  \caption{Results obtained from classification analysis from from student and non-student populations}
  \label{tab:results_students}
  \begin{tabular}{p{1.55cm}|p{1.45cm}|p{0.85cm}|p{0.85cm}|p{0.85cm}|p{0.85cm}|p{0.85cm}|p{0.85cm}|p{0.85cm}|p{0.85cm}|p{0.85cm}|p{0.85cm}}
    \toprule
    Population & Method & \multicolumn{10}{c}{Personality Trait}\\
    \cline{3-12}
     & & \multicolumn{2}{c|}{Extraversion} & \multicolumn{2}{c|}{Agreeableness} & \multicolumn{2}{c|}{Consc.} & \multicolumn{2}{c|}{Neuroticism} & \multicolumn{2}{c}{Openness}\\
     \cline{3-12}
     & & Acc (\%) & $\kappa$ & Acc (\%) & $\kappa$ & Acc (\%) & $\kappa$ & Acc (\%) & $\kappa$ & Acc (\%) & $\kappa$\\
     \midrule
     Student & Method 2 & 73 & 0.45 & 71 & 0.43 & 75 & 0.51 & 78 & 0.56 & 76 & 0.52\\
     \cline{1-12}
     \rule{0pt}{3ex}Non-Student & Method 2 & 69 & 0.37 & 74 & 0.47 & 72 & 0.43 & 75 & 0.51 & 73 & 0.45\\
  \bottomrule
\end{tabular}
\end{table}

Our work here facilitates the development of trait prediction models by providing the first evidence that traits can be reliably predicted from smartphone sensor data originating from a range of distinct countries. Future work may aim to extend this contribution by replicating these results across a broader range of countries, in a more high-powered study and with age groups beyond 44 years. \textcolor{black}{Moreover, smartphone penetration in Africa, Asia and Oceania is increasing every year\footnote{https://www.statista.com/chart/17148/smartphone-adoption-by-world-region/}\footnote{https://www.statista.com/statistics/257048/smartphone-user-penetration-in-india/} -- further research in automatic personality detection should bring insights from outside of the Euro-American continents. This would not only provide a better understanding of cross-cultural differences in personality models but it would also bring attention and give another perspective on automatic behavior analysis outside Western-centric settings}.

\section{Implications} 

The relevance of personality traits in the technology industry is growing rapidly. Companies have traditionally used persuasive appeals to encourage individuals to purchase a specific product, however, recent research suggest that these persuasive appeals are more effective in changing behavior when they are targeted to the individuals' psychological traits. For instance, Matz et al. targeted individuals' psychological characteristics to tailor persuasive appeals by predicting their personality traits from their digital footprints~\cite{matz2017psychological}. Their research suggests that tailoring messages to individual personality traits has the potential to significantly increase the effectiveness of persuasive marketing leading to an increase in conversion rate of advertisers. 

Personality traits have also been linked to health behaviors~\cite{raynor2009associations} and perceptions of wellbeing~\cite{cloninger2011personality}. Recent work by Khwaja et al. looked at improving a person's subjective wellbeing by providing personalised activity recommendations that are aligned with their personality traits~\cite{khwaja2019recsys}. Halko et al. showed a range of relationships between personality and persuasive technologies, aiming to personalize and improve health-promoting applications~\cite{halko2010personality}. 

Hence, predicting personality traits from sensing data provides immense potential for technology service providers to increase the user experience of their average consumer, and provide effective health intervention solutions. Our results have direct implications on the practical application and replicability of smartphone based personality models, and ultimately on adapting smartphone experience to different kinds of users.

\section{Conclusion}
This paper is the first to show a systematic link between diverse set of features (both situational and personal) and personality trait prediction models in a multi-country sample. Our study demonstrated that developing country-agnostic personality models (applicable in new countries) is feasible with classification accuracy of up to 71\%; however significant improvements, up to 7\%, can be achieved by considering culture-specific datasets. \textcolor{black}{Furthermore, we showed that the use of a gender-balanced, gender-independent and age-balanced datasets can improve prediction accuracy by up to 17\%.}

Our findings highlight the importance of Noise, Location, Unlock events and Accelerometer data as the most predictive data categories for personality trait models across different countries. In addition to exploring similarities, we also unpacked differences in predictive features for specific Big Five Traits across countries. \textcolor{black}{Normalized frequency distributions of top predictive features (especially those obtained from Noise and Unlock events data) across different countries showed that there are behavioural similarities observed between Spain and UK, and between Colombia, Peru and Chile. This indicates that there are continent-wise model similarities in Europe versus South America.}

Lastly, we observed a high proportion (up to 30\%) of users who opted out of providing access to Location, Noise, and Call data, even for research purposes. Considering the predictive power of these data categories for personality models, this is an important aspect to take into consideration both by model developers as well as service providers.

\section*{Acknowledgment}
This work has been supported by the European Union's Horizon 2020 research and innovation programme, under the Marie Sklodowska-Curie grant agreement no. 722561.

\bibliographystyle{ACM-Reference-Format}

\end{document}